\documentclass[twocolumn,prb,showpacs,preprintnumbers,amsmath,amssymb]{revtex4}

\usepackage{graphicx}

\begin{document}


\title{Underlying Fermi surface
 of Sr$_{14-x}$Ca$_x$Cu$_{24}$O$_{41}$ in two-dimensional momentum space observed by angle-resolved photoemission spectroscopy}

\author{T. Yoshida$^1$, X. J. Zhou$^2$,
Z. Hussain$^3$, Z.-X. Shen$^4$, A. Fujimori$^1$, H. Eisaki$^5$, S.
Uchida$^1$} \affiliation{$^1$Department of Physics, University of
Tokyo, Bunkyo-ku, Tokyo 113-0033, Japan} \affiliation{$^2$National
Laboratory for Superconductivity, Beijing National Laboratory for
Condensed Matter Physics, Institute of Physics, Chinese Academy of
Sciences, Beijing 100080, China} \affiliation{$^3$Advanced Light
Source, Lawrence Berkeley National Lab, Berkeley, CA 94720, USA}
\affiliation{$^4$Department of Applied Physics and Stanford
Synchrotron Radiation Laboratory, Stanford University, Stanford,
CA94305, USA} \affiliation{$^5$National Institute of Advanced
Industrial Science and Technology, Tsukuba 305-8568, Japan}
\date{\today}

\begin{abstract}
We have performed an angle-resolved photoemission study of the
two-leg ladder system Sr$_{14-x}$Ca$_x$Cu$_{24}$O$_{41}$ with $x$=
0 and 11. ``Underlying Fermi surfaces" determined from low energy
spectral weight mapping indicates the quasi-one dimensional nature
of the electronic structure. Energy gap caused by the charge
density wave has been observed for $x$=0 and the gap tends to
close with Ca substitution. The absence of a quasi-particle peak
even in $x$=11 is in contrast to the two-dimensional high-$T_c$
cuprates, implying strong carrier localization related to the hole
crystalization.
\end{abstract}


\maketitle Since the discovery of the high-$T_c$ superconductor in
layered cuprates, one-dimensional (1D) cuprates have also
attracted much interest \cite{DagottoReview}. Particularly, Cu-O
ladders with an even number of coupled Cu-O chains are predicted
to have a finite spin gap reminiscent of that in the underdoped
high-$T_c$ cuprates, and have the possibility of superconductivity
with hole doping \cite{Dagotto}. Ladder compounds such as
Sr$_{14-x}$Ca$_x$Cu$_{24}$O$_{41}$ (Sr14-24-41) and LaCuO$_{2.5}$
in which two-dimensional CuO$_2$ planes is reorganized into 1D
segments have long been envisioned to bridge 1D chain systems,
where Luttinger liquid behaviors are predicted, and 2D plane
systems, which is the stage of the high-$T_c$ superconductivity.
Sr14-24-41 is composed of alternating stacks of the plane of edge
sharing CuO$_2$ chains, (Sr, Ca) layer, and the plane including
two-leg Cu$_2$O$_3$ ladders. With Ca substitution, holes are
transferred from the chain sites to the ladder sites
\cite{Osafune}. From the optical reflectivity measurement of
Sr14-24-41 at room temperature, the number of holes doped into the
Cu$_2$O$_3$ ladder plane is already 0.07/Cu atom for $x$=0, which
should be enough to realize superconductivity (SC) in 2D cuprate
\cite{Osafune}. However, the $x$=0 samples are insulating and show
an activated behavior in the resistivity \cite{Motoyama}. Upon Ca
substitution for Sr, holes localized in the chains are transferred
to the ladders and induce mobile carriers on the ladder. When $x>$
11, the mobile carriers exhibit superconductivity under high
pressure \cite{Uehara}. To understand the electronic structure of
such quasi-1D systems, angle-resolved photoemission spectroscopy
(ARPES) is a powerful tool because it provides us with
momentum-resolved information about the electronic states. In the
previous ARPES experiments on the ladder system Sr14-24-41, two
dispersive features along the ladder and chain directions were
observed and assigned to the ladder-derived band (near $E_F$) and
the chain-derived band ($\sim$ 1 eV below $E_F$), respectively
\cite{Takahashi, Sato}. However, in the previous results, only
one-dimensional momentum dependence along the ladder/chain
direction has been studied in momentum space. Since
superconductivity in this system appears when application of
pressure causes a dimensional crossover from one to two as
reflected in electrical resistivity \cite{Nagata}, two-dimensional
electronic structure has been thought to be crucial to the
superconductivity. In the present work, we have performed ARPES
experiments on the two-leg ladder compound Sr14-24-41 with $x$=0
and 11 and clarified spectral weight distribution in the
two-dimensional momentum space. The ARPES results clearly
demonstrate the quasi-one-dimensional electronic structure of the
ladder but with finite energy dispersion perpendicular to the
ladder direction. Also, an energy gap of the ladder band of
$\sim70$ meV observed for $x$=0 tends to close with Ca
substitution. The line shape of the gapped states will be compared
with those for the 2D high-$T_c$ cuprates.

The ARPES measurements were carried out at BL10.0.1 of Advanced
Light Source, using incident photons of 55.5 eV. We used a SCIENTA
SES-200 analyzer with total energy resolution of 20 meV and
momentum resolution of 0.02$\pi/c$, where $c$ = 3.95 \textrm{\AA}
is the Cu-O-Cu distance along the ladder direction. The lattice
constant perpendicular to the ladder within the ladder plane is
$a$= 11.46 \textrm{\AA}. We studied high quality single crystals
of Sr$_{14-x}$Ca$_x$Cu$_{24}$O$_{41}$ with $x$= 0 and 11 grown by
the traveling-solvent floating-zone method. Measurements were
performed in an ultra high vacuum of 10$^{-11}$ Torr. The samples
were cleaved \textit{in situ} and measured at 20 and 150 K for the
$x$=11 and 0 samples, respectively.  In the present measurements,
the electric field vector $\mathbf{E}$ of the incident photons lie
in the ladder and chain plane.

\begin{figure}
\includegraphics[width=8cm]{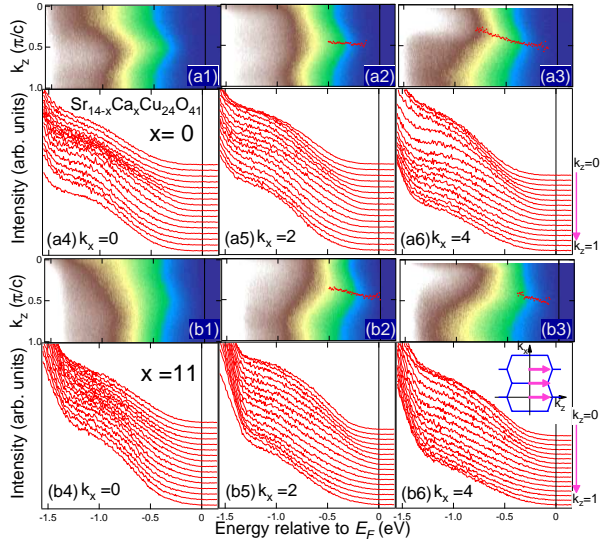}
\caption{\label{EkEDC} (Color online) ARPES spectra for
Sr$_{14-x}$Ca$_x$Cu$_{24}$O$_{41}$ ($x$= 0, 11). Panels (a1)-(a3)
and (b1)-(b3) show spectral intensities in energy-momentum
($E$-$k$) space along the $k_z$ direction for $k_x$($\pi$/a)= 0, 2
and 4. Red dots are the peak position of the MDC's. Panels
(a4)-(a6) and (b4)-(b6) show energy distribution curves (EDC's)
corresponding to the $E$-$k$ intensity map in (a1)-(a3) and
(b1)-(b3), respectively. Inset for panel (b6) illustrates
corresponding cuts in momentum space and the ladder Brillouin
zone.}
\end{figure}

Figure \ref{EkEDC} shows spectral intensity in energy-momentum
($E$-$k$) space along the ladder ($k_z$) direction and
corresponding energy distribution curves (EDC's) for these $k_x$
values. Red dots in the $E$-$k$ maps are the peak position of the
momentum distribution curves (MDC's). Here, MDC's reflect energy
dispersion of the ``quasi-particle" although the MDC width is very
broad. In panels (a4) and (b4), a broad structure around -1 eV
below the Fermi level ($E_F$) is assigned to energy bands from the
chain states according to the previous studies \cite{Takahashi,
Sato}. On the other hand, the structure around -0.5 eV at
$k_z\sim$ 0.5 can be ascribed to the ladder electronic structure,
since the dispersion at $|k_z| <$ 0.5 well correspond to that
predicted by band-structure calculation \cite{Arai}. This
structure shows clear dispersion and have a bottom of the band
around $k_z$=0 as shown in panels (a6) and (b6). Although MDC's
show the remnant of a dispersive band, a sharp quasi-particle peak
in the EDC's was not observed even for the $x$=11 sample. The
intensity of the ladder component is enhanced in the second
Brillouin zone (BZ) [Fig. \ref{EkEDC}(a6) and (b6)], while that of
the chain band is suppressed. Since the ladder can be regarded as
a portion of the CuO$_2$ plane, the enhancement of the ladder
spectra in the second BZ may be caused by similar matrix-element
effects as seen in the high-$T_c$ cuprates. The spectral
intensities of the chain bands are enhanced in the first BZ
[panels (a4) and (b4)], while those in the second BZ is strongly
suppressed [panels (a6) and (b6)]. These contrasted behaviors of
the spectral intensity between the ladder and the chain bands
manifest the difference in the orbital symmetry of the wave
functions, since the Cu 3$d_{xy}$ orbitals for the chain band are
rotated by 45 degree from the Cu 3$d_{x^2-y^2}$ orbitals for the
ladder band.

\begin{figure}
\includegraphics[width=8cm]{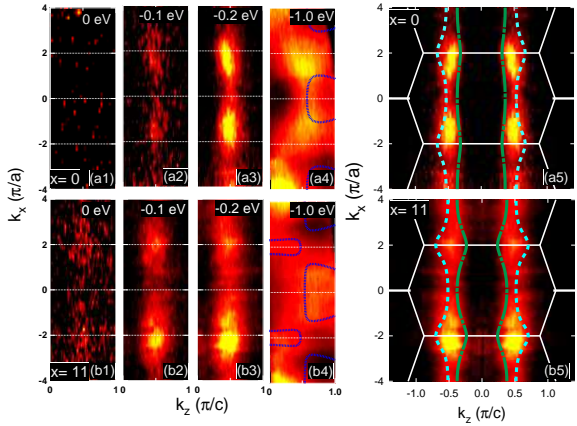}
\caption{\label{nk}(Color online) Spectral weight mapping of the
ARPES spectra of Sr$_{14-x}$Ca$_x$Cu$_{24}$O$_{41}$ ($x$= 0, 11).
Panels (a1)-(a4) and (b1)-(b4) show the intensity map for $x$=0
and 11 at each energy from the $E_F$, respectively. Dotted lines
in panels (a4) and (b4) indicate the Fermi surfaces of the chain
band predicted by band-structure calculation \cite{Arai}. Panels
(a5) and (b5) are intensity maps for -0.2 eV symmetrized with
respect to the $k_z$=0 line and represent the ``underlying Fermi
surfaces". Dashed and dash-dotted lines indicate Fermi surfaces of
the bonding and anti-bonding bands of the ladder predicted by the
band-structure calculation \cite{Arai}.}
\end{figure}

Spectral weight at various binding energies are mapped in momentum
space in Fig. \ref{nk}. From comparison between $x$=0 and 11, the
low energy spectral weight for $x$= 11 is more intense than that
in $x$=0 [Fig. \ref{nk} (a2) and (b2)]. Particularly, the map at
$E_F$ for $x$=0 shows almost no intensity, indicating that the
energy gap is opened on the entire "Fermi surface". While the low
energy spectra represent the quasi-one-dimensional Fermi surface
shape of the electronic structure of the ladder, in the
high-energy range $\sim$ -1 eV [Fig. \ref{nk} (a4) and (b4)], the
spectral weight distribution is more widely distributed in
momentum space and is more strongly $k_x$ dependent, i.e.,
two-dimensional. These structures come from the chain states and
are similar to the Fermi surface of chain state predicted by the
band-structure calculation \cite{Arai} as superimposed on the
mapping in panels (a4) and (b4). However, note that the chain
states is not observed near the Fermi level but observed in the
high-energy range $\sim$ 1 eV, since the holes in the chain is
strongly localized unlike those in the ladders.

In Fig. \ref{nk}(a5) and (b5), the observed spectral weight at the
energy of -0.2 eV are symmetrized with respect to the $k_z$=0 line
in two-dimensional momentum space. One can see that the spectral
weight distribution is approximately confined between the bonding
and anti-bonding Fermi surfaces predicted by the band calculation
\cite{Arai}. Furthermore, the intensity around the zone boundary
is enhanced in both samples. The observed intensity modulation
along the $k_x$ direction indicates finite inter-ladder hopping
integrals, which cause the bonding and anti-bonding band
splitting, although the EDC's are too broad to separate the
bonding and anti-bonding bands. The spectral intensity of $x$=0 in
the first BZ is stronger than those in the second BZ, while $x$=11
shows opposite behavior. This difference may come from the
intensity of the chain strucure in the high energy region. As
shown in Fig. \ref{EkEDC}(a4) and (b4), the chain structure in
$x$=0 is clearer than that in $x$=11. The observed change in the
spectral weight distribution by Ca substitution indicate spectral
weight transfer from the high energy to the low energy in the
ladder electronic states, which is related to the hole transfer
from the chain to the ladder \cite{Osafune}.

\begin{figure}
\includegraphics[width=8cm]{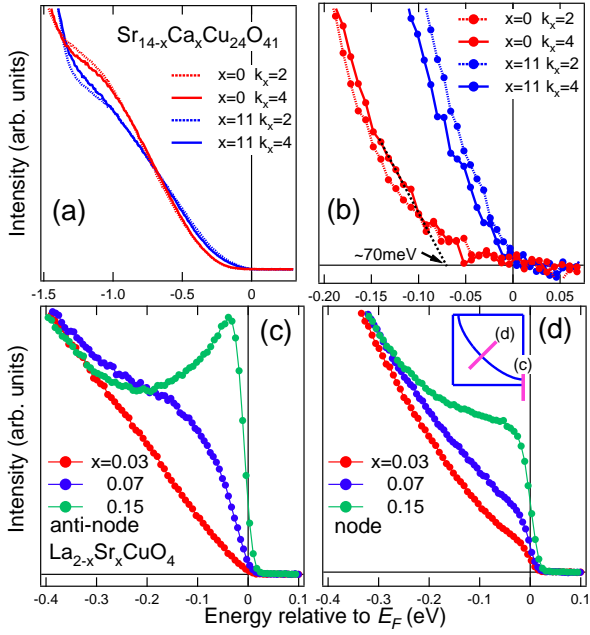}
\caption{\label{IMDC}(Color online) Integrated intensities of MDC
as a function of energy for Sr$_{14-x}$Ca$_x$Cu$_{24}$O$_{41}$.
(a) Integrated spectra along the $k_z$-axis direction. Panel (b)
is an enlarged plot of panel (a). Gap size for $x$=0 are obtained
as illustrated by a black dotted line. Panels (c) and (d) show
integrated ARPES spectra of two-dimensional high-$T_c$ cuprates
La$_{2-x}$Sr$_x$CuO$_4$ along ($\pi$,0)-($\pi$,$\pi$) and the
(0,0)-($\pi$,$\pi$) directions, respectively, (see inset) for
comparison with the present results. Note that the hole doping
level of the ladder is $\sim$0.07 for $x$=0 and $\sim$0.2 for
$x$=11 \cite{Osafune}.}
\end{figure}

We have estimated the energy gap near $E_F$ from the integration
of ARPES spectra on a cut along the $k_z$ direction as shown in
Fig. \ref{IMDC}. As shown in the Fig. \ref{IMDC}(a) and (b), the
energy gap size for $x$=0 can be estimated as $\Delta\sim$ 70 meV,
consistent with the result of the optical conductivity,
$2\Delta\sim$ 130 meV. The optical gap is interpreted as a charge
density wave (CDW) gap \cite{Vuletic}. For the $x$=11 sample,
although the slope of the spectra reach $E_F$, no clear Fermi edge
is observed. Also, the integrated MDC spectra for $k_x$=2 is
slightly closer to the $E_F$ than that in $k_x$=4, while there is
almost no difference between them for $x$=0. This can be taken as
a signature of the increased two-dimensionality in $x$=11.
However, the change in $x$=11 is still small and far from a
two-dimensional electronic structure. The transport properties in
the ladder direction of $x$=11 show metallic behavior
($d\rho/dT>$0) \cite{Nagata} and a Drude peak in the optical
conductivity \cite{Osafune} at high temperatures. However, the
electrical resistivity shows localization behavior at low
temperatures and the Drude peak has a suppression in the low
energy ($\sim$ 10meV) region. Therefore, we conclude that the
broad spectral line shape without clear Fermi edge reflects the
localization at low temperatures seen in the transport properties.

The absence of QP is contrasted with the case of the lightly doped
high-$T_c$ cuprates, which shows clear a QP in the nodal
direction, although the hole concentration $\sim$ 0.2 of the
ladder for the $x$=11 sample is almost the same as that for the
overdoped cuprates \cite{Osafune}. In order to compare the present
results with the high-$T_c$ cuprates in Fig.\ref{IMDC}(c) and (d),
we show integrated ARPES spectra of La$_{2-x}$Sr$_x$CuO$_4$ (LSCO)
\cite{YoshidaPRB,YoshidaJPCM} along the ($\pi$,0)-($\pi$,$\pi$)
and (0,0)-($\pi$,$\pi$) directions, respectively. In LSCO, because
of the pseudogap around ($\pi$,0), the intensity decrease with
approaching $E_F$. Particularly, the spectral line shape of
$x$=0.03 is similar to the ladder spectra for $x$=11 shown in
Fig.\ref{IMDC}(b). Since the shape of the Fermi surface around
($\pi$,0) in the underdoped 2D cuprates is a quasi-one-dimensional
\cite{Zhou}, this similality in the spectral line shape implies
that  the mechanism of the pseudogap formation is similar between
the ladder and the 2D cuprates.  On the other hand, the clear
Fermi edge indicating the $E_F$ crossing of QP in the nodal region
of LSCO [Fig.\ref{IMDC} (d)] is contrasted with the case of the
ladder, probably reflecting the difference between the 1D and 2D
electronic structures. This 1D-2D difference would be related to
the facts that two-dimensionality makes hole carriers mobile and
that three-leg ladder compounds can easily become metallic with
hole doping compared to the two-leg ladder system. In a two-leg
ladder, Zhang-Rice singlet has a strong tendency toward hole
crystalization as observed in the X-ray scattering even in $x$=11
samples \cite{Abbamonte}.

Finally, let us discuss the condition for the occurrence of
superconductivity in the ladder compounds. We have found that the
electronic structure is still quasi-one dimensional even in $x$=11
samples. On the other hand, when the superconductivity occurs
under high pressure, the transport properties become rather
two-dimensional, i.e., anisotropy of the $a$- and $c$- axis
resistivities become small \cite{Nagata}. As for the electronic
structure, the relatively isotropic transport properties may be a
result of a topological change of the Fermi surface from one
dimension to two dimension. The shift of the chemical potential
caused by the hole transfer from the chain to the ladder under
high pressure may cause a two-dimensional Fermi surface of the
ladder bands. Therefore, two-dimensional electronic structure may
be necessary to the superconductivity analogous to the
two-dimensional high-$T_c$ cuprates.

In summary, we have performed an ARPES study of the two-leg ladder
system Sr14-24-41 to investigate the Ca substitution effects on
the two-dimensional electronic structure. The intensity modulation
due to inter-ladder hopping has been observed, indicating finite
2D effect. The CDW gap size for $x$=0 is about 70 meV and reduced
to 10-20 meV for $x$=11, consistent with the results of the
optical conductivity \cite{Vuletic}. We did not, however, observe
a clear QP peak even in the metallic $x$=11 samples. This behavior
is contrasted with that of the lightly doped LSCO, which shows a
clear QP peak crossing $E_F$ in the nodal direction. Possible
origins of the absence of a QP are attributed to the strong
localization of doped holes due to the quasi-one dimensionality.

This work was supported by a Grant-in-Aid for Scientific Research
in Priority Area ``Invention of Anomalous Quantum Materials",
Grant-in-Aid for Young Scientists from the Ministry of Education,
Science, Culture, Sports and Technology and the U.S.D.O.E.
contract DE-FG03-01ER45876 and DE-AC03-76SF00098. ALS is operated
by the Department of Energy's Office of Basic Energy Science,
Division of Materials Science.

\bibliography{LadderBib}

\end{document}